# DySkew: Dynamic Data Redistribution for Skew-Resilient Snowpark UDF Execution

Chenwei Xie, Urjeet Shrestha, Corbin McElhanney, Lukas Lorimer, Gopal V*,
Zihao Ye, Yi Pan, Nic Crouch, Elliott Brossard, Florian Funke, Yuxiong He

*Snowflake, Inc*

*Abstract*— Snowflake revolutionized data warehousing with an elastic architecture that decouples compute and storage, enabling scalable solutions for diverse data analytics needs. Building on this foundation, Snowflake has advanced its AI Data Cloud vision by introducing Snowpark, a managed turnkey solution that supports data engineering and AI/ML workloads using Python and other programming languages. While Snowpark's User-Defined Function (UDF) execution model offers high throughput, it is highly vulnerable to performance degradation from data skew, where uneven data partitioning causes straggler tasks and unpredictable latency. The non-uniform computational cost of arbitrary user code further exacerbates this classic challenge. This paper presents DySkew, a novel, data-skew-aware execution strategy for Snowpark UDFs. Built upon Snowflake's new generalized skew handling solution, an adaptive data distribution mechanism utilizing per-link state machines. DySkew addresses the unique challenges of user-defined logic with goals of fine-grained per-row mitigation, dynamic runtime adaptation, and low-overhead, cost-aware redistribution. Specifically, for Snowpark, we introduce crucial optimizations, including an eager redistribution strategy and a Row Size Model to dynamically manage overhead for extremely large rows. This dynamic approach replaces the limitations of the previous static round-robin method. We detail the architecture of this framework and showcase its effectiveness through performance evaluations and real-world case studies, demonstrating significant improvements in the execution time and resource utilization for large-scale Snowpark UDF workloads.

## I. Introduction

Modern data platforms increasingly rely on flexible, programmable computation frameworks to support complex analytics, data engineering, and machine learning workloads. Snowflake's Snowpark [1] framework extends the Snowflake [2] AI Data Cloud with a unified API that allows developers to express custom logic in languages such as Python, Java, and Scala, running in Snowpark sandbox [3] to benefit from the platform's native performance, scalability, and governance. Instead of pushing data out to external compute, Snowpark brings computation into Snowflake's execution environment, enabling UDF, Stored Procedures, and DataFrame-style transformations to run close to the data with consistent security and operational guarantees. By integrating custom code execution with Snowflake's elastic, distributed processing engine, Snowpark offers a powerful abstraction for developing advanced data applications. However, this tight coupling of user-defined logic with Snowflake's distributed execution model introduces new performance challenges—particularly around data skew, where uneven input partitioning can lead to straggler tasks, underutilized compute, and unpredictable latency.

While Snowpark provides a seamless programming model, its execution of UDFs inherits the classic challenges of distributed data processing [4]. In Snowflake's execution model, data is partitioned and scheduled across multiple compute nodes, and UDF invocations are executed in parallel on these partitions. This approach delivers high throughput for embarrassingly parallel workloads, but it also exposes performance sensitivity to the distribution of input data. When data is well-balanced, Snowpark UDFs achieve near-linear scalability as additional compute resources are provisioned. However, real-world datasets—such as event logs, customer telemetry, semi-structured records, and operational tables—often exhibit significant skew along key dimensions [5]. Even modest skew can cause severe runtime imbalance: some partitions complete quickly while others become stragglers, prolonging the overall job and causing resource underutilization or unnecessary scaling. Moreover, because Snowpark UDFs encapsulate arbitrary developer-defined logic, their computational cost is often opaque to the optimizer, making traditional cost-based balancing strategies insufficient.

Addressing these challenges requires a data-skew–aware execution strategy specifically designed for the characteristics of Snowpark UDF workloads. Unlike traditional relational operators, UDF performance can vary dramatically at the per-row level, and the system must respond to this variability without incurring excessive overhead. To that end, our design is guided by three key goals:

- Fine-grained skew handling: UDFs may involve computationally intensive per-row logic, meaning even a single heavy record can dominate task runtime. Mitigation must therefore operate at a finer granularity at per-row level.

- Dynamic runtime adaptation: Because different rows may trigger highly variable execution costs, static or precomputed partitioning is insufficient. The system should observe worker latency in real time and continuously rebalance work based on observed performance.

- Low-overhead redistribution: Skew mitigation must avoid excessive data movement or over-reactive rebalancing. Redistribution decisions should be cost-aware so that the overhead of transferring rows does not exceed the performance gains from improved balance.

This paper presents the Snowflake solution, DySkew, to handle generalized data skew issues in Snowpark UDF execution, detailing its essential components. We first recognize the challenges imposed by the Snowpark execution model and present the limitations of the previous Snowpark skew handling logic, which redistributed rows across Python interpreter processes using a round-robin approach. We further explain the challenges specific to Snowpark and the special handling implemented on top of the generalized data skew handling solution. Next, we present the benchmark results and performance improvement from real production cases to show how DySkew brings significant performance advantages to customers.

## II. Background - Snowpark UDF

Snowpark is a set of libraries and code execution environments that enable customers to run workloads in Python and other programming languages next to their data in Snowflake. Furthermore, Snowpark provides code execution environments where customers' arbitrary Python code can be pushed down and executed in Snowpark's secure sandbox, which is the infrastructure to isolate customers' Python workloads from Snowflake's SQL data warehouse workloads. To address the performance bottleneck of Python global interpreter lock, Snowpark also launches multiple Python interpreter processes to take advantage of multiple CPU cores in each worker when running custom UDFs, which can process the data in parallel. This multiple-Python-interpreter setup increases parallelism on a per-node basis. However, if data is unevenly distributed, it can lead to imbalanced workloads, preventing us from achieving optimal performance.

In this section, we will first explain the overall Python function execution model and then current data skew handling logic as well as its limitations.

### A. Python Function Execution

Python Functions are executed as part of Snowflake SQL queries. At query startup time, depending on users' Python code, necessary Python packages are installed on the Snowflake virtual warehouse nodes. Since Python has a global interpreter lock, Snowpark creates many Python interpreter processes for each function in the query. Snowpark initializes the Python interpreter before forking additional processes to reduce initialization time. The Virtual Warehouse worker threads communicate with Snowpark Python interpreter processes through gRPC [7] to pass Rowsets [8] for computation. Data goes through a set of transformations between Virtual Warehouse worker threads and Snowpark Python Interpreters within the same virtual warehouse node.

### B. Maintaining the Integrity of the Specifications

Snowpark's architecture setup can introduce data skew issues when unevenly distributed workloads are assigned to worker processes, and lead to performance degradation. This is especially severe for Snowpark workload because user code may take a longer time to process a single row. For example, assume we have 4 rows and 4 python interpreters, if each row needs to take 1 minute to process, it would take 4 minutes if all 4 rows land into the same Python interpreters while it only takes 1 minute if the rows are evenly distributed and each Python interpreter can handle one row.

To resolve this issue, we introduced an optimization to redistribute the rows. During the query compiler time, we will insert a special link between Snowpark operators and its source producer, during the execution stage, the source producer will see the special link and redistribute the rows across all Python interpreter processes in different virtual warehouse nodes using a round robin approach [6], as shown in Fig 1, ensuring full parallelism.

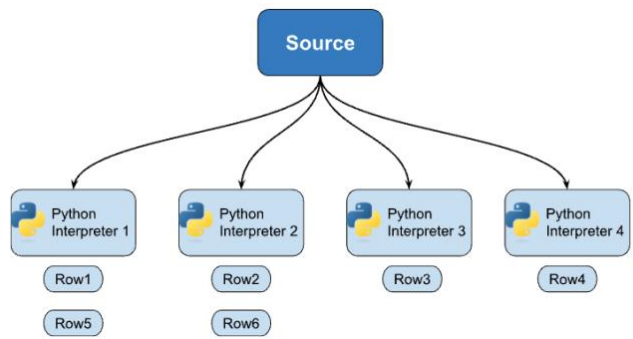

Fig. 1. Round Robin Per Row Redistribution

This solution shows improvements both on industry standard benchmark like TPCX-BB [9] and real customer workloads, however, we also notice a few limitations with this approach:

**Suboptimal distribution:** The redistribution is static by the nature of round robin approach, the rows will be evenly distributed to each worker [10], However, an even distribution of rows is not optimal because different rows can have significantly different processing times due to variations in data complexity or the specific logic within the UDF. The computational capacity of individual workers may vary since workers may also be simultaneously handling other unrelated workloads.

**Conflict with Data Locality:** The nature of the round-robin approach forces data to be evenly distributed across all available workers, including those that may be on remote nodes or outside the immediate cluster processing the query. This universal redistribution can break the fundamental assumption of data locality for other parts of the query execution plan. In specific, complex query scenarios where data locality is critical for correctness or performance [11], applying this skew handling solution may lead to data corruption or incorrect query results. Therefore, this solution cannot be safely applied to all Snowpark UDF use cases.

**Solution Divergence and Maintenance Burden:** Data skew is a general problem across all of Snowflake's processing engines. As noted, the engineering team is actively developing a generalized, platform-wide skew handling logic specifically for SQL workloads. Implementing and maintaining a completely separate, bespoke skew-handling solution dedicated only to Snowpark UDFs results in duplicated efforts and increased maintenance efforts [12].

**Complicated threshold:** Distributing the rows incurs overhead, necessitating a special threshold to determine if redistribution is beneficial. To calculate this, we build a historical metrics tracking system to store the per-row execution time for each User-Defined Function, significantly increasing the complexity of the whole system. Furthermore, a single, static threshold exhibits overfitting and fails to generalize effectively across diverse UDF execution scenarios [13].

### III. Dynamic Redistribution

Data skew can happen across the whole stage of a query execution. From the beginning of the data scan, where the data is unevenly distributed across different partitions, to each of the execution stages, where the data might redistribute because of the query characteristics, eg, Topk, order [14]. As a high performance analysis product, we need to fix skew issues wherever they exist so we can reduce single query runtimes by improving CPU utilization and minimize the performance impact when the skew is present. In the following section, we will first explain the generalized skew handling solution in Snowflake and Snowpark specific optimizations because of the extra challenges.

#### A. Generalized Skew Handling

Each stage of Snowflake query execution is represented by an operator. Data flows between operators through data links that connect a producer operator to a consumer operator. The default data link maintains a 1:1 mapping between producer and consumer instances, each producer instance sends data exclusively to its corresponding consumer instance. This design implies that data skew originating in the producer will persist through the consumer and propagate along the entire execution pipeline.

The solution we propose is replacing the default data link with an adaptive data link. This new link employs an adaptive data distribution mechanism that uses independent state machines for each link instance to dynamically decide how to route data between producer and consumer operators. The key insight is that data distribution decisions are made per-link-instance rather than globally, allowing fine-grained adaptation to runtime conditions. Furthermore, each state machine can observe the state of its sibling instances to make informed redistribution decisions.

**State machine:** For each instance of the new data link, it's an independent state machine [15], Fig 2 represents an example of one possible state machine. By default, the state machine will follow the red links, the other transitions are supported based on different redistribution policies, which is explained in the following section.

The state machine follows a general progression through four phases:

- Phrase1: Initial State. The initial state represents the link before data processing begins. During this phase, the link is configured with its redistribution policy and prepares to receive data. The initial state determines the link's starting behavior based on the redistribution policy.

- Phrase2: Deciding State. In the deciding state, the link actively processes data while continuously evaluating runtime conditions through the skew model. Based on the skew model's assessment, the state machine can transition to either continue local processing or begin redistribution to remote workers. The deciding phase represents the adaptive core of the system.

- Phrase 3: Intermediate State. Intermediate states handle special conditions during transitions. For example, when transitioning from local to distributed processing, the link may need to complete in-flight file boundaries before fully switching modes. Intermediate states ensure correctness during transitions without violating semantic guarantees.

- Phrase 4: Terminal State. Terminal states represent final operating modes where the redistribution decision has been committed. In non-looping configurations, once a terminal state is reached, the link will not change its behavior for the remainder of the query.

**Redistribution Policy:** The redistribution policy is a preference declared by the consumer operator that governs whether and when redistribution is permitted. Different redistribution policies select different transition tables within the state machine, determining which states are reachable:

- Never Distribute. The link must never send rows to remote workers. All data is processed locally regardless of skew conditions. This policy is required for operators with ordering requirements or local state dependencies.

- Distribute Late. The link begins by processing data locally and only redistributes if the skew model detects significant imbalance. This is the default policy, optimizing for locality while maintaining the ability to adapt.

- Distribute Early. The link begins redistributing immediately, assuming distributed processing is beneficial. This policy is typically used for cases where computation cost is unpredictable and parallelization is generally advantageous.

Skew Detection Models: The state machine and redistribution policies provide the mechanism for handling data skew, but a critical question remains: how do we determine if skew exists at runtime? We developed multiple skew detection models that observe runtime metrics across sibling instances to answer this question.

All models employ an N-strikes framework: skew must be detected N consecutive times before triggering redistribution. This reduces sensitivity to transient fluctuations and avoids false positives.

- Row percentage-based model. This model compares the row count of the current instance against the average row count of its sibling instances. If the current instance has processed significantly more rows than the average (exceeding a threshold percentage), it is considered skewed. Formally, instance $i$ is skewed if the follow matches, where $R_i$ is the row count of instance $i$, $R_{-i}$ is the average row count of other instances:

$$R_i \times \theta > R_{-i} \quad (1)$$

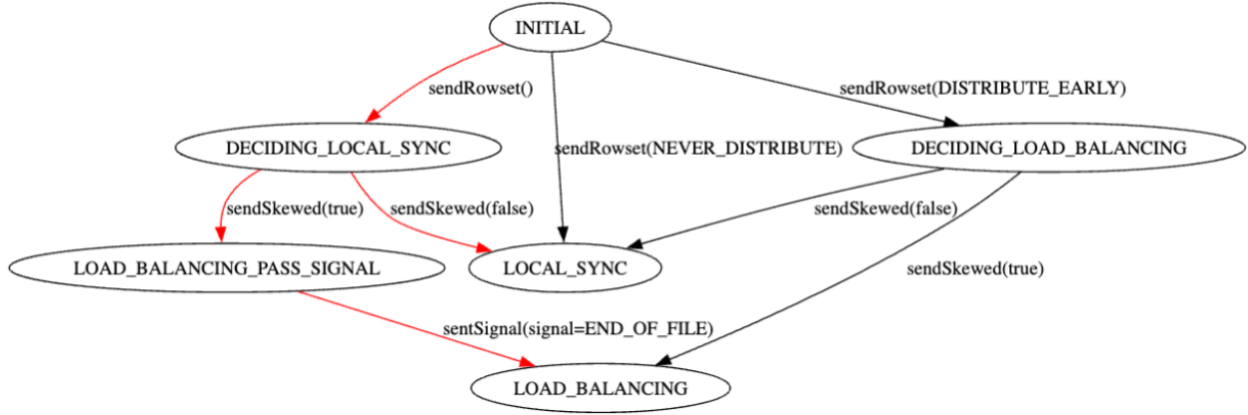

Fig. 2. State machine for Generalized Skew Handling

- Idle time based model. This model detects skew by observing whether sibling instances are sitting idle while the current instance is busy. An instance is considered idle if it has not received a row or signal for a configurable period. If the number of idle siblings exceeds a threshold, the current instance is considered skewed and should redistribute work to utilize those idle resources. This model directly measures resource underutilization rather than inferring it from row counts, making it effective for workloads with variable per-row processing costs such as UDFs.

- Synchronous Time Spent Model. This model tracks the cumulative time each instance spends inside the consumer operator processing rows synchronously. Rather than comparing absolute values, the model computes the rate of change of sync time over a sliding window of measurements. If the current instance's sync time is growing significantly faster than the average of its siblings, it indicates accelerating workload imbalance, where $\frac{DS_i}{dt}$ is the slope of sync time for instance $i$.

$$\frac{DS_i}{dt} \times \theta \geq \frac{DS_{-i}}{dt} \quad (2)$$

*B. Snowpark Optimizations*

The Generalized Skew Handling solution explained in the above section works well for the SQL workload, however, while we adopt this for Snowpark workloads, we discovered that Snowpark user code has a different set of assumptions than regular SQL code. Thus, we further improve the solution with the consideration of those specific characteristics for Snowpark workloads.

**Snowpark Workloads Are Highly Sensitive to Skew:** Snowpark workloads exhibit a heightened sensitivity to data skew compared to native SQL operations. While Snowflake's C++ engine processes rows with minimal overhead, Snowpark UDFs execute within managed Python or Java environments. These runtimes are inherently slower than native code and require mandatory data serialization across process boundaries [16]. This architectural disparity creates a skew amplification effect: a marginal skew that would be negligible in C++ can result in a >50% increase in total UDF latency. Consequently, standard skew detection thresholds are often too insensitive to capture the disproportionate impact of straggling workers on Snowpark workloads.

Benchmarking revealed that for UDFs, the performance gain from redistribution almost always outweighs the cost of data movement. We implemented an Eager Redistribution policy within the state machine. Unlike the "Distribute Late" default, this policy bypasses the initial observation phase and begins redistributing rows immediately, ensuring maximum utilization of available worker processes from the start of the operator's lifecycle.

**Redistribution Overhead Scales with Row Size:** Although eager redistribution is generally beneficial, it can cause regressions when the rows are large. In scenarios where UDFs process extremely large objects (e.g., high-resolution images or massive JSON blobs), the network and serialization overhead of moving a single row can exceed the time saved by remote execution [17], especially on workloads without significant skew. In edge cases, we observed performance regressions of up to 20x due to redistributing 100GB+ of data unnecessarily.

We solved this by introducing a heuristic that monitors batch density. While Snowflake typically targets thousands of rows per batch, this density drops by over 99% when processing large objects. Our model evaluates two concurrent signals: the Idle Time Model (for skew) and the Batch Density (for row size). If the operator is not skewed and the batch density falls below a certain number of rows per batch, the state machine triggers a transition to the Terminal State, disabling redistribution. This intervention prevents network saturation from "heavy rows" while maintaining high-throughput parallelism for the vast majority of standard Snowpark workloads.

**Forced Remote Distribution:** The generalized skew-handling framework includes a mechanism that forces the redistribution logic to prioritize remote workers over the local instance to improve skew detection sensitivity. However, for Snowpark workloads already employing an Eager Redistribution policy, this "self-skipping" behavior is counterproductive. Forcing a worker to bypass itself leaves local CPU cycles idle and incurs unnecessary network costs to move data that could have been processed locally. Benchmarking showed this caused significant regressions compared to a location-agnostic strategy, especially when running on a smaller amount of nodes.

We optimized the Snowpark execution path by removing the mandatory self-skipping logic. Since the system assumes an imbalanced state from the outset, the discovery benefit of avoiding the local instance is obsolete. By allowing the producer to treat the local worker as a valid destination, the system ensures local compute resources are fully saturated alongside remote workers. This eliminates self-exclusion bias, reduces total network traffic, and ensures that eager parallelization does not result in local underutilization.

## IV. Performance evaluations

To evaluate DySkew's scalability, approximately 150 customer queries were replayed across different cluster sizes to compare the legacy static strategy against the new adaptive mechanism. The results indicate that DySkew's performance gains correlate directly with virtual warehouse size. As the number of compute nodes increases, data skew becomes more probable and impactful, making dynamic redistribution more critical. While there is a slight regression at 2 nodes, 4-node and 8-node configurations showed significant latency reductions, highlighted by a nearly 10% improvement in P99 tail latency.

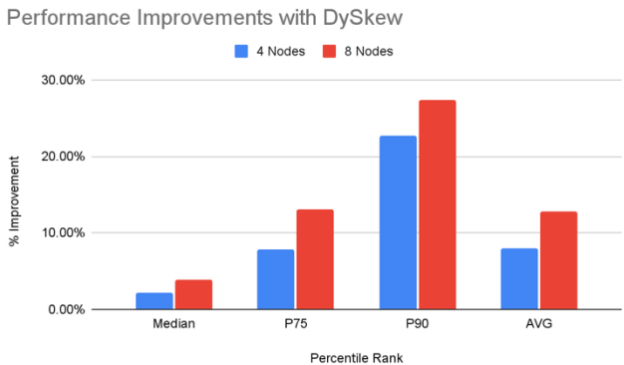

Fig. 3. Performance improvements with DySkew on Customer queries

Using the TPCx-BB benchmark on 4 nodes, our analysis of queries incorporating UDFs demonstrates significant improvements for Query 10 and Query 19, which saw performance improvements of 43% and 36% respectively. The other queries had no significant differences (<5%).

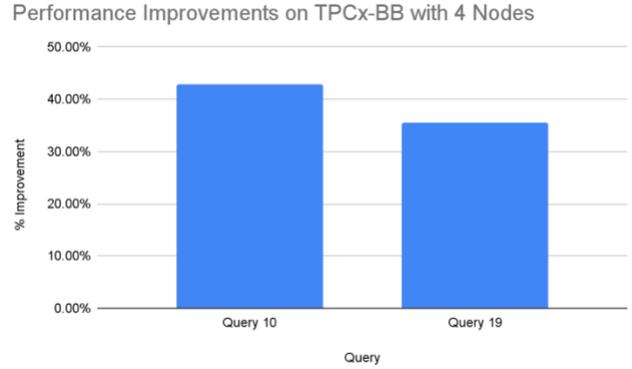

Fig. 4. Performance improvements with DySkew on TPCx-BB

Lastly, we monitored the deployment of DySkew across production Snowpark environments. Our telemetry indicates that redistribution is now automatically applied to 37.6% of all Snowpark UDF queries. Following the rollout, we observed a steady downward trend in the P99 execution time for these workloads, resulting in a total performance improvement of 20.4%.

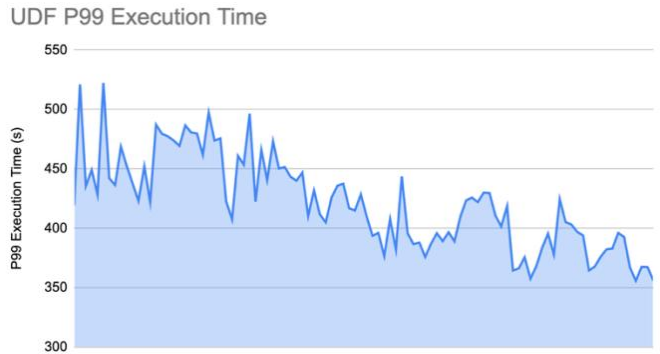

Fig. 5. Performance improvements with DySkew on producation

ACKNOWLEDGMENT

DySkew is the result of a coordinated, cross-functional effort involving the SQL, Snowpark, Performance Engineering, and Customer Support teams. We extend our deepest gratitude to the numerous talented individuals across these organizations whose hard work and dedication were essential to developing this dynamic redistribution framework. We are especially thankful to the partner teams for their incredible support and their passion for bringing this innovative, skew-resilient solution to our users. It is an honor and a privilege to work with such an outstanding, multi-disciplinary team, and we are constantly inspired by their collective excellence.

REFERENCES

[1] B. Baker et al., "Snowpark: Performant, Secure, User-Friendly Data Engineering and AI/ML Next To Your Data," in 2025 IEEE 45th International Conference on Distributed Computing Systems Workshops (ICDCSW), Glasgow, United Kingdom, 2025, pp. 213-218, doi: 10.1109/ICDCSW63273.2025.00042.

[2] B. Dageville, T. Cruanes, M. Zukowski, V. Antonov, A. Avanes, J. Bock, J. Claybaugh, D. Engovatov, M. Hentschel, J. Huang, A. W. Lee, A. Motivala, A. Q. Munir, S. Pelley, P. Povinec, G. Rahn, S. Triantafyllis,

and P. Unterbrunner. "The Snowflake Elastic Data Warehouse." In Proc. of ACM SIGMOD, 2016.

[3] Gaurav Jain et al., "SEE++: Evolving Snowpark Execution Environment for Modern Workloads". arXiv preprint arXiv:2511.12457.

[4] J. Wang, D. Crawl, S. Purawat, M. Nguyen and I. Altintas, "Big data provenance: Challenges, state of the art and opportunities," 2015 IEEE International Conference on Big Data (Big Data), Santa Clara, CA, USA, 2015, pp. 2509-2516, doi: 10.1109/BigData.2015.7364047.

[5] Y. Le, J. Liu, F. Ergün and D. Wang, "Online load balancing for MapReduce with skewed data input," IEEE INFOCOM 2014 - IEEE Conference on Computer Communications, Toronto, ON, Canada, 2014, pp. 2004-2012, doi: 10.1109/INFOCOM.2014.6848141.

[6] Creating Automated Optimizations for Python User-Defined Functions with Snowpark's Parallel Execution - https://www.snowflake.com/en/engineering-blog/snowpark-parallel-python-udf-optimization/

[7] gRPC - A High-Performance, Open-Source Universal RPC Framework. https://www.grpc.io/.

[8] R. Kimball, and M. Ross. "The Data Warehouse Toolkit: The Definitive. Guide to Dimensional Modeling. 3rd ed." Hoboken, NJ: John Wiley & Sons, 2013.

[9] TPCx-BB - A Big Data Benchmark. https://www.tpc.org/tpcx-bb/default5.asp

[10] Devi, D. C. (2016). Load balancing in cloud computing environment using improved weighted round robin algorithm for nonpreemptive dependent tasks. The Scientific World Journal, 2016, 1–14. https://doi.org/10.1155/2016/3896065

[11] Giceva, J., et al. (2014). "Deployment of Query Plans on Multicores." Proceedings of the VLDB Endowment (PVLD), 8(3), pp. 233-244.

[12] Sloan, J. J. (1990). "Software Complexity and Software Maintenance Costs." MIT Thesis Archive.

[13] Borkar, V. R., Carey, M. J., & Li, C. (2012). "Big data platforms: What's next?" XRDS: Crossroads, The ACM Magazine for Students.

[14] Schneider, F. B. (1990). "Implementing Fault-Tolerant Services Using the State Machine Approach: A Tutorial." ACM Computing Surveys, 22(4), pp. 299-319.

[15] Saur, K., et al. (2022). "Containerized execution of UDFs: An experimental evaluation." Proceedings of the VLDB Endowment (PVLDB), 15(11).

[16] Saur, K., et al. (2022). "Containerized execution of UDFs: An experimental evaluation." Proceedings of the VLDB Endowment (PVLDB), 15(11)

[17] Li, S., Hu, S., & Li, J. (2015). "A Survey of Data Skew Handling in MapReduce." International Journal of Parallel Programming, 43(3).